\documentclass[a4paper,draft,12pt]{article}
\usepackage{amssymb}

\usepackage{graphicx}
\usepackage{amsmath}
\usepackage{a4}

\input{tcilatex}

\begin{document}

\centerline{\bf Incomplete descriptions and relevant entropies}\medskip

\centerline{\bf Roger Balian}\medskip

\centerline{Service de Physique Th\'eorique, CEA/Saclay}

\centerline{F-91191 Gif-sur-Yvette Cedex, France}

\bigskip

%TCIMACRO{
%\TeXButton{noindent}{\noindent%
%}}%
%BeginExpansion
\noindent%
%
%EndExpansion
\textbf{Abstract. }Statistical mechanics relies on the complete though
probabilistic description of a system in terms of all the microscopic
variables. Its object is to derive therefrom static and dynamic properties
involving some reduced set of variables. The elimination of the irrelevant
variables is guided by the maximum entropy criterion, which produces the
probability law carrying the least amount of information compatible with the
relevant variables. This defines relevant entropies which measure the
missing information (the disorder) associated with the sole variables
retained in an incomplete description. Relevant entropies depend not only on
the state of the system but also on the coarseness of its reduced
description. Their use sheds light on questions such as the Second Law, both
in equilibrium an in irreversible thermodynamics, the projection method of
statistical mechanics, Boltzmann's \textit{H}-theorem or spin-echo
experiment.

\section{\textbf{Introduction}}

The concept of entropy underlies both thermodynamics and statistical
physics, but its subtleties and its multiple aspects make it difficult to
grasp. Our initial motivation is to understand how the entropy of
thermodynamics $S_{\mathrm{th}}$ is related to the entropy of statistical
mechanics. The solution of this problem will turn out to require the
introduction in statistical physics of several different entropies, which
will enlighten some other questions.

For systems at \textit{equilibrium, }the situation is simple, and it is
dealt with in many tutorial books. At the macroscopic scale, the state of a
system at equilibrium is characterized by some set of thermostatic
variables, such as the particle number, the energy and the volume for a
fluid. (We distinguish the thermodynamics of equilibrium from the
thermodynamics of irreversible processes by terming them as ``\textit{%
thermostatics}'' and ``\textit{thermodynamics}'' proper, respectively.) The
entropy $S_{\mathrm{th}}$ is some function of these variables, a function
that can be determined indirectly by experiments. At the microscopic scale,
the same state is described by a density operator $D$ having the
Boltzmann--Gibbs exponential form. The \textit{von Neumann entropy}
associated to $D$, defined by

\begin{equation}
S(D)\equiv -k\,\text{\textrm{Tr}}D\,\text{\textrm{ln\thinspace }}D, 
\tag{1.1}
\end{equation}
is then identified with the \textit{thermostatic entropy} $S_{\mathrm{th}};$
the Boltzmann constant $k=1.38\times 10^{-23}$ JK$^{-1}$entering (1.1)
accounts for the conventional choice of the kelvin as the unit of
temperature in thermal physics. This identification is justified through
derivation of the Laws of thermostatics --- in particular the Second Law ---
from the Boltzmann--Gibbs distributions of statistical physics. Moreover the
microscopic approach allows us in principle to evaluate $S_{\mathrm{th}},$
which macroscopically is an empirical quantity.

The identification of the entropy $S_{\mathrm{th}}$ of thermodynamics with
the von Neumann entropy $S(D)$ cannot, however, be extended to \textit{%
dynamical} processes. Consider an isolated system off equilibrium. On the
one hand, according to the Second Law, its thermodynamic entropy $S_{\mathrm{%
th}}$ is larger in its final equilibrium state than in its initial state. If
its evolution proceeds through states close to equilibrium, the
Clausius--Duhem inequality expresses moreover that the time-derivative of $%
S_{\mathrm{th}}$ is non-negative. On the other hand, the microscopic
evolution is governed (in the Schr\"{o}dinger picture) by the \textit{%
Liouville--von Neumann equation} for the density operator,

\begin{equation}
\text{i}\hslash \frac{dD}{dt}=\left[ H,D\right] ,  \tag{1.2}
\end{equation}
where $H$ is the hamiltonian operator of the system. It follows from (1.1)
and (1.2) that $S(D)$ does not vary with time. This constancy, when compared
to the Second Law, constitutes a modern form of the \textit{paradox of
irreversibility} (W. Thomson 1874, J. Loschmidt 1876), which expresses a
qualitative difference between microscopic and macroscopic dynamics. The
solution of this paradox requires of course the thermodynamic entropy $S_{%
\mathrm{th}}$ of non-equilibrium states to differ from the von Neumann
entropy $S\left( D\right) .$

Actually, we shall associate below (\S 6 and \S 7) various so-called
relevant entropies with a given density operator $D,$ and shall show (\S 10)
that one of them can be identified with $S_{\mathrm{th}}.$ We shall more
generally illustrate the utility of the idea of relevant entropy, by showing
how it underlies the projection method (\S 8 and \S 9) and how it enlightens
some questions of irreversibility in statistical physics (\S 11 and \S 12).
As a preliminary step, we briefly review (\S \S 2 to 5) some aspects of the
concepts of state and of entropy in statistical physics, on which we shall
rely in the following. In particular it will be important to distinguish the
complete statistical description of a state (\S 2) from its incomplete
descriptions (\S 5) and to grasp the meaning of the maximum entropy
criterion (\S 4). Although this article is self-contained, it is somewhat
sketchy. Further explanations, details and developments can be found in the
first few and last few chapters of reference $[1].$

\section{Complete statistical description: the density operator}

The state of a system at a given time can be defined in several ways that
should not be confused.

(i) \textit{Thermodynamics} relies on the identification of some set of 
\textit{macroscopic quantities} $\mathcal{A}_{i},$ the control of which is
sufficient to govern the whole macroscopic behaviour of the system. A state
of this system is then characterized by specifying the quantities $\mathcal{A%
}_{i},$ irrespective of the microscopic variables which otherwise remain
arbitrary. At the microscopic level, the thermodynamic description is thus 
\textit{incomplete}.

(ii) At the other extreme, one can imagine states, termed as ``\textit{%
microstates}'', in which all the variables are defined inasmuch as possible.
In classical mechanics, when the uncertainly principle may be disregarded,
nothing prevents such a definition of the variables to be perfect: a
microstate is characterized by specifying the positions and momenta of all
the particles (within permutations in the labels of indistinguishable
particles). In quantum mechanics, a microstate must retain probabilistic
features; it is represented as a ket, a vector in the Hilbert space
associated with the system.

(iii) For large systems it is however impossible in practice to prepare a
microstate. It is thus necessary to resort to \textit{statistics}. The
definition of a state therefore refers, not to a single system, but to a
typical system chosen out of a statistical \textit{ensemble} of systems all
prepared under similar conditions. We recall in the continuation of this
section the main features of this last type of state.

The quantum \textit{physical quantities}, or ``\textit{observables}'', are
represented as hermitean \textit{operators} $A$ which act in the Hilbert
space of the system. In the classical limit, an observable tends to a 
\textit{random variable} which is a function of the (random) positions and
momenta of the particles. Thus, in quantum mechanics, the observables play
the r\^{o}le of non-commutative random variables.

We characterize statistically a state by its \textit{density operator }at
the considered time. This means that we can evaluate the expectation value $%
\left\langle A\right\rangle $ of any observable $A$ at that time by means of
the equation

\begin{equation}
\left\langle A\right\rangle =\mathrm{Tr}AD.  \tag{2.1}
\end{equation}
The description of the system furnished by its density operator $D$ is thus 
\textit{statistical}, since in general the variance $\left\langle
A^{2}\right\rangle -\left\langle A\right\rangle ^{2}$ does not vanish, and 
\textit{complete}, since any expectation value can be derived from (2.1).

The density operator $D$ is therefore the tool which implements the \textit{%
correspondence} $A\mapsto \left\langle A\right\rangle $ from the whole set
of observables $A$ to their expectation values $\left\langle A\right\rangle
. $ Actually, in classical statistical mechanics, $D$ is replaced by an
ordinary probability distribution, from which the expectation value of any
observable results through an integration which replaces the trace of (2.1).
Thus a density operator \ should be regarded as a probability distribution
for non-commuting random variables. Conversely, a linear correspondence $%
A\mapsto \left\langle A\right\rangle $ defines a density operator $D.$ The
standard properties of $D,$ normalization, hermiticity and positivity,
express that the expectation value of the unit operator $I$ is $\left\langle
I\right\rangle =1,$ and that $\left\langle A\right\rangle $ is real, $%
\left\langle A^{2}\right\rangle $ is positive for any hermitean $A.$

A density operator $D$ is usually represented as a\textit{\ matrix }$%
D_{\alpha \beta }$ in the \textit{Hilbert space.} However, it may be
convenient (and it will be necessary in \S 8) to regard the two indices $%
\alpha ,\beta $ of this density matrix as a single, compound index. The
density operator is then represented, no longer as a matrix but as a \textit{%
vector} in the so-called \textit{Liouville space}, where the compound index $%
\alpha ,\beta $ plays the r\^{o}le of a single set of coordinates. We can
then perform linear transformations on the compound index. This defines 
\textit{Liouville representations }of quantum statistical mechanics, which
encompass the standard matrix representations but are more general and more
flexible \cite{balazs}\cite{hillery}\cite{balian2}\cite{rau}. In a
Liouville representation, the observables are also represented as vectors
with a compound index. They belong to the \textit{dual Liouville space,} and
an expectation value (2.1), written as

\begin{equation}
\left\langle A\right\rangle =\underset{_{\alpha \beta }}{\sum }A_{\alpha
\beta }D_{\beta \alpha }=\text{\textbf{(}}A\,\text{\textbf{;}}D\text{\textbf{%
)},}  \tag{2.2}
\end{equation}
appears as a \textit{scalar product }\ between the two vectors $A$ and $D$
of the two conjugate Liouville spaces. In a Liouville representation, the
Liouville--von Neumann equation, which governs the evolution of $D,$ takes
the form

\begin{equation}
\frac{dD}{dt}=\mathsf{L}D,  \tag{2.3}
\end{equation}
which is equivalent to (1.2). The liouvillian $\mathsf{L}$ is a \textit{%
superoperator}, represented in the Liouville space as a matrix. In the
Hilbert space, it appears as a tensor with $2\times 2$ indices, and (2.3) is
written in terms of the density matrix $D_{\alpha \beta }$ as

\begin{equation}
\frac{dD_{\alpha \beta }}{dt}=\underset{_{\gamma \beta }}{\sum \,}\mathsf{L}%
_{\alpha \beta ,\gamma \delta }D_{\delta \gamma };  \tag{2.4}
\end{equation}
comparison with the more standard notation (1.2) shows that

\begin{equation}
\mathsf{L}_{\alpha \beta ,\gamma \delta }=\frac{1}{\text{i}\hslash }\left(
H_{\alpha \delta }\delta _{\beta \gamma }-H_{\gamma \beta }\delta _{\alpha
\delta }\right) .  \tag{2.5}
\end{equation}

The simplest example of Liouville representation concerns the statistical
state of a spin $\frac{1}{2}.$ Rather than describing it by a $2\times 2$
hermitean density matrix with trace 1, it is handy to represent it by a
vector in a 3-dimensional Liouville space, with merely the expectation
values of the Pauli matrices as components. The equation of \ motion (2.3)
then describes directly the Larmor precession of this vector in a plane
perpendicular to the magnetic field.

Another example of Liouville representation, the Wigner representation, is
useful in the study of the classical limit of quantum statistical mechanics.
In this limit, the observables tend to commuting random variables. For a
system of $N$ indistinguishable particles, the density operator $D$ in the
Wigner representation tends to an ordinary probability distribution in the 6$%
N$-dimensional phase space of coordinates and momenta of these particles; it
can be identified with the probability density in phase space of classical
statistical mechanics. The traces in (1.1) and in (2.1) or equivalently the
scalar product (2.2) reduce to integrals in phase space, with the measure

\begin{equation}
\underset{i=1}{\overset{N}{\prod }}d^{3}\mathbf{r}_{i}d^{3}\mathbf{p}%
_{i}/N!\,h^{3N}  \tag{2.6}
\end{equation}
which arises from the summation over $\alpha $ and $\beta $ of (2.2). The
equation of a motion (2.3) directly yields the classical Liouville equation.

\section{The statistical entropy as missing information}

Information theory associates with an event $n,$ which has the probability $%
p_{n}$ to occur, its \textit{surprisal} $I_{n}=-k\,\ln p_{n},$ a number
which measures the amount of information that we gain when we get to know
the occurrence of this event. Within the multiplicative constant $k$ which
defines the unit of information, this logarithmic expression for $I_{n}$ is
imposed by the condition that the information is additive when knowledge is
gained by steps. The surprisal vanishes when the occurrence of $n$ is
certain, it increases as its probability becomes weaker. We shall let $k=1$
in the continuation. This will imply that Boltzmann's constant is replaced
by 1, so that we take the joule as the unit of temperature.

For a set of exclusive events $n,$ one of which is expected to occur, the 
\textit{statistical entropy} is then defined as the average amount of
information which is gained when either event occurs, that is,

\begin{equation}
S\left( \left\{ p_{n}\right\} \right) =\underset{n}{\sum \,}p_{n}I_{n}=-%
\underset{n}{\sum \,}p_{n}\ln p_{n}.  \tag{3.1}
\end{equation}
This expression, a function of the probabilities $p_{n}$, equivalently
measures the amount of information which is \textit{missing }when only the
probabilities $p_{n}$ of occurrence of the various events $n$ are known.

Returning to statistical mechanics, the eigenvalues $p_{n}$ of a density
operator $D$ may be interpreted as ordinary probabilities of the microstates
which are the corresponding eigenvectors of $D$ and which behave as
exclusive events. The expression (3.1), rewritten as

\begin{equation}
S\left( D\right) =-\mathrm{Tr}D\ln D,  \tag{3.2}
\end{equation}
is identified with the \textit{von Neumann entropy} (1.1). It is thus
interpreted as the \textit{lack of information }which arises from the
incompleteness of our statistical description by means of the density
operator $D.$ Like the density operator itself, the von Neumann entropy is a
statistical quantity, since it is a natural \textit{measure of uncertainty.}
As such, it is not a property of the object at hand, but has partly a
subjective nature linked to our knowledge of this object, which is extracted
from a statistical ensemble.

This subjective character is somewhat hidden in the alternative
interpretation of $S\left( D\right) $ as a \textit{measure of disorder. }We
can argue, however, that ``disorder'' and ``missing information'' are
synonymous. Consider, for instance, the 13 spades taken from a pack of
cards. A first configuration, in which they are set according to the
decreasing values of bridge, ace, king, ..., two, displays perfect order;
its statistical entropy vanishes. Some second configuration, reached after a
long shuffling, appears completely disordered. Nothing allows us to
recognize any special feature in it. The reason for this difference is that
we regard the former configuration as unique (with probability 1), while we
regard the latter as just one among all 13! possible configurations. Before
looking at it, we assign to it the probability of occurrence 1/13!, and the
entropy is therefore ln13!. However, if the shuffling is performed by a
skilful conjurer who controls every card, he knows perfectly the arrangement
in the second configuration, which \textit{for him }is perfectly ordered.
Indeed he can reshuffle it back to the first configuration.

A similar situation occurs in \textit{spin-echo experiments.} At the initial
time, the spins are prepared in a perfectly ordered configuration: they all
point in the $x$-direction. A permanent magnetic field is applied in the $z$%
-direction, but its magnitude is not quite the same at the site of each
spin, so that the Larmor frequencies of the various spins are slightly
dispersed around their average value. From the time 0 on, the spins precess
rapidly in the $xy$-plane, but at slightly different speeds, so that the
total magnetization $M$ decreases in length while rotating at the average
Larmor frequency: it \textit{relaxes to zero.} We thus reach after some time 
$T$ a completely disordered configuration where the spins point towards
arbitrary directions of the $xy$-plane. However, this configuration is not
disordered for the spin-echo experimentalist, who knows where it was issued
from. He takes advantage of this knowledge by applying to the system at the
time $T$ a brief pulse of magnetic field along $x,$ which suddenly rotates
each spin by $\pi ,$ replacing it by its symmetric with respect to the $x$%
-axis. Thus, the spins which were ahead in the precession are now behind and
conversely. After a second delay $T,$ the precession, which again takes
place at different speeds according to the site of each spin, produces back
the original ordered configuration. We shall return in \S 12 to this example
which suggests that order, or entropy, is not defined uniquely in the
seemingly disordered configuration attained after relaxation.

\section{The maximum entropy criterion}

Owing to its interpretation as the missing amount of information associated
with the probability distribution $p_{n}$, the statistical entropy (3.1) is
currently used as a tool for \textit{statistical inference.} The purpose of
statistical inference is to make reasonable predictions about some
quantities, starting from an exact or a statistical knowledge of some other
quantities. We thus have to assign to the considered set of elementary
events $n$ a probability distribution $p_{n},$ which should of course
account for the available data but should otherwise be unbiased.

It is natural to admit that, among two probability laws, the \textit{more
biased }one is the one which carries \textit{more information.} This
assumption leads us to assign the probabilities $p_{n}$ of each event by
relying on the maximum entropy criterion, as was first advocated by Jaynes $%
[6]$. The knowledge of the available data, whether they are specified
exactly or statistically, first sets up constraints on the probabilities $%
p_{n}.$ Then, among all probability laws compatible with these constraints,
we select the one for which the \textit{statistical entropy} (3.1) \textit{%
is largest.} Any other probability law compatible with the data would have a
smaller statistical entropy, i.e., would carry more information than the
minimum amount needed to account for these data. Choosing probabilities by
means of the maximum entropy criterion thus amounts to retain all the
available information while \textit{discarding any other irrelevant
information. }

This method has successfully been applied to numerous and varied problems,
ranging from signal theory to image processing and from detection of
astrophysical objects to determination of protein structures. Its extension
to statistical mechanics relies on the maximization of the von Neumann
entropy (3.2), a function of $D$ which measures the lack of information. A
density operator is thereby assigned to a state which is characterized by
some given conditions, by looking for the maximum of $S\left( D\right) $
under constraints that account for these conditions. We shall work out this
procedure in \S 6 and \S 7 after having shown (\S 5) some examples of such
constraints. A direct justification of the maximum entropy criterion in
statistical mechanics, based on Laplace's indifference principle (equal
probabilities should be assigned when nothing in known) rather than on the
existence of von Neumann's entropy, is given in ref. $[7]$.

\section{Relevant observables and incomplete descriptions}

We have noted (\S 2,i) that the state of a system is characterized at the
macroscopic level by some partial set of quantities $\mathcal{A}_{i}.$ When
only such an information is available, the microscopic description of the
state is incomplete, since we know nothing about the other microscopic
variables. We shall identify a macroscopic quantity $\mathcal{A}_{i}$ with
the expectation value $\left\langle A_{i}\right\rangle $ of the
corresponding observable $A_{i}$. We thus deal with a system for which 
\textit{only the expectation values} $\left\langle A_{i}\right\rangle $ of
some set of observables $A_{i},$ that we refer to as the relevant set $%
R\equiv \left\{ A_{i}\right\} ,$ are specified. Such a partial knowledge,
both statistical and incomplete, is not sufficient to characterize the
density operator $D,$ which is merely constrained to satisfy the set of
equations

\begin{equation}
\mathrm{Tr}\text{ }DA_{i}=\left\langle A_{i}\right\rangle =\mathcal{A}_{i}. 
\tag{5.1}
\end{equation}

Note that for some other quantities the correspondence from macrophysics to
microphysics is more straightforward. For instance, if the particle number
of a system is specified \textit{exactly,} leaving no room for any
statistical fluctuation, this particle number enters the microscopic
description through the very definition of the Hilbert space of the system.
As another example, the volume $\Omega $ of a fluid appears in the
microscopic theory implicitly through the hamiltonian. Such quantities are
accounted for directly in the microscopic theory, and not as expectation
values (5.1) of observables.

Let us illustrate the above correspondence between macroscopic quantities $%
\mathcal{A}_{i}$ and relevant observables $A_{i}$ with a few examples.

In \textit{thermostatics,} the variables $\mathcal{A}_{i}$ are those which
characterize the macroscopic state in equilibrium. For instance the
thermostatic state of a simple fluid, made of a single type of identical
molecules enclosed in a vessel with volume $\Omega ,$ is characterized by
two macroscopic data $\mathcal{A}_{i},$ namely its particle number $\mathcal{%
N}$ and its internal energy $\mathcal{U}$. We wish to describe this system
at the microscopic level in terms of a density operator $D$ in the Fock
space, that is, in the Hilbert space with an arbitrary number of particles.
Our first datum is then the expectation value $\left\langle N\right\rangle =%
\mathcal{N}$ of the particle number operator (in the Fock space), while the
internal energy $\mathcal{U}$ is identified with the expectation value $%
\left\langle H\right\rangle $ of the hamiltonian operator $H.$ The set $R$
of relevant observables contains here two elements, the operators $N$ and $H$%
. We noted above that the volume $\Omega $ enters the problem through the
expression of $H,$ which contains a potential confining the particles within 
$\Omega .$

Our identification of macroscopic data $\mathcal{A}_{i}$ with microscopic
expectation values $\left\langle A_{i}\right\rangle $ raises a question,
since the microscopic quantities, particle number and energy, are allowed to 
\textit{fluctuate }freely around their expectation values, in contrast to
their macroscopic counterparts. However, the statistical fluctuations which
appear in statistical mechanics for macroscopic systems turn out to be
extremely small (of order $\left\langle N\right\rangle ^{-1/2}$ in relative
value); they are actually much smaller than the experimental uncertainties
on the variables $\mathcal{A}_{i},$ so that this qualitative discrepancy
between the macroscopic and the microscopic descriptions is ineffective.

For other systems in thermostatic equilibrium, the set $R$ of relevant
observables $A_{i}$ can involve, apart from $N$ and $H,$ additional \textit{%
constants of the motion,} represented by observables which commute with $H.$
When some invariance is broken, we have moreover to include \textit{order
parameters,} which also keep a fixed value in time at least on the
time-scales of observation.

For \textit{thermodynamic processes} close to equilibrium, a state is
described macroscopically at each time by distinguishing within the system a
set of \textit{subsystems }$\mathit{a}$, each of which is nearly in local
equilibrium $[8]$. The macroscopic state is then characterized by specifying
the thermostatic variables for each subsystem. For instance, in the
thermodynamic or hydrodynamic regime, a simple fluid can be analyzed into
volume elements, larger than the mean free path, so as to have reached local
equilibrium, but sufficiently small so as to be practically homogeneous. The
variables $\mathcal{A}_{ia}$ are then the \textit{number of particles }in
each volume element, their total \textit{energy }and their total \textit{%
momentum }(or equivalently the local densities of particles, energy and
momentum, smoothed over the mean free path); the compound index $ia$ denotes
here both the location $a$ in space and the nature $i$ of the physical
quantity. The relevant set $R$ of observables $A_{ia}$ are the corresponding
operators (or random variables in the classical limit).

A similar analysis holds quite generally, for \textit{thermal, mechanical,
chemical, electrical }or \textit{magnetic }processes, provided they are
sufficiently slow so as to ensure local equilibrium. The thermodynamic
quantities are most often the locally conserved quantities, but may also
include other variables varying slowly on the microscopic scale.

For more rapid processes, these thermodynamic variables are not sufficient
to determine the physics of the system; more detailed quantities should be
included into the set $\mathcal{A}_{i},$ although many microscopic variables
are still kept aside. For instance, the description of a gas suited for the 
\textit{ballistic regime, }or \textit{Boltzmann description,} requires an
analysis, not only in the ordinary space, but in the 6-dimensional \textit{%
single-particle phase space.} The set of relevant variables $\mathcal{A}_{i}$
are then the densities $f\left( \mathbf{r},\mathbf{p}\right) $ of particles
at each point of this space; the index $i$ stands here for $\left( \mathbf{%
r,p}\right) .$ In the Boltzmann description, all the 2-point, 3-point, ...
correlations are disregarded. At the microscopic level, the set $R$ of
relevant observables includes all the single-particle observables, but only
these. The components of the density operator (or, in the classical limit,
of the probability density in the 6$N$-dimensional phase space) referring to
2-particle, 3-particle... observables remain unspecified.

Other examples of incomplete descriptions are provided, on the one hand by 
\textit{coarse-graining,} on the other hand by \textit{collective variables.}
Consider for instance \textit{large nuclei.} It is impossible to describe
their state fully, but their dynamics can be accounted for by following only
some set of collective variables, such as the ones which describe the shape
of the nucleus in a fission problem.

The set of relevant observables always includes the quantities which can
experimentally be observed or controlled. However, the above examples show
that for dynamical problems our description also includes other quantities,
needed to account theoretically for the physics of the system. For instance,
in fluid dynamics, the densities of particles, energy and momentum are not
controlled everywhere, but they must be introduced to write the basic,
Navier--Stokes and Fourier, equations.

\section{Least biased states and relevant entropies}

We return to inference in statistical mechanics, first focusing on the
following \textit{single-time problem.} Let $R\equiv \left\{ A_{i}\right\} $
be some set of relevant observables. We assume that their expectation values 
$\left\langle A_{i}\right\rangle =\mathcal{A}_{i}$ are specified at some
given time, but that nothing else is known. From this incomplete
information, we wish to make a reasonable statistical guess about the
expectation value $\left\langle B\right\rangle $ of any other observable $B$
at the considered time. Equivalently, we wish to construct at that time the
density operator $D$ which encompasses our whole statistical knowledge (\S
2), both within and outside $R.$

To this aim, we use the maximum entropy criterion of \S 4. The density
operator $D_{0}$ to be assigned to the system should both fulfil the \textit{%
constraints} (5.1) expressing that $\left\langle A_{i}\right\rangle =%
\mathcal{A}_{i}$ and render the von Neumann entropy (3.2)\textit{\ maximum,}
so as to avoid including in $D$ any additional information. The construction
of the least biased state $D_{0}$ is therefrom achieved through introduction
of a \textit{lagrangian multiplier} $\gamma _{i}$ associated with each
constraint (5.1). The result has the form

\begin{equation}
D_{0}=\exp \left[ -\,\Psi -\underset{i}{\sum }\gamma _{i}A_{i}\right] , 
\tag{6.1}
\end{equation}
where the term $\Psi $ accounts for the normalization Tr$D_{0}=1$ of $D_{0}$%
. The exponential in (6.1) reflects the occurrence of a logarithm in $%
S\left( D\right) .$

A direct proof of (6.1) relies on the inequality

\begin{equation}
S\left( D\right) <-\mathrm{Tr}D\ln D^{\prime },  \tag{6.2}
\end{equation}
which holds for any pair of density operators $D$ and $D^{\prime }$. If in
(6.2) we replace $D^{\prime }$ by (6.1) and $D$ by any other density
operator satisfying the constraints (5.1), the right-hand side reads

\begin{equation}
-\mathrm{Tr}D\ln D_{0}=\Psi +\underset{i}{\sum }\gamma _{i}\left\langle
A_{i}\right\rangle =-\mathrm{Tr}D_{0}\ln D_{0}=S\left( D_{0}\right) , 
\tag{6.3}
\end{equation}
and we readily see that $S\left( D_{0}\right) >S\left( D\right) ,$ as we
anticipated.

The determination of $D_{0}$ is achieved by specifying the values of the
parameters $\Psi $ and $\gamma _{i}$. The normalization condition, written as

\begin{equation}
\Psi \left( \left\{ \gamma _{i}\right\} \right) =\ln \mathrm{Tr}\text{%
\thinspace }\exp \left[ -\underset{i}{\sum }\gamma _{i}A_{i}\right] , 
\tag{6.4}
\end{equation}
first defines $\Psi $ as a function of the multipliers $\gamma _{i}$. The
latter quantities are then determined by inserting (6.1) into (5.1), which
yields a set of equations,

\begin{equation}
-\frac{\partial \Psi }{\partial \gamma _{i}}=\mathcal{A}_{i},  \tag{6.5}
\end{equation}
relating the multipliers $\gamma _{i}$ and the macroscopic data $%
\left\langle A_{i}\right\rangle =\mathcal{A}_{i}.$

The above procedure is standard in \textit{thermostatics. }If for instance
the relevant set $\left\{ A_{i}\right\} $ consists of the particle number
operator $N$ and the hamiltonian $H$, we find thus a statistical
justification for the grand canonical density operator \ (6.1). The data $%
\mathcal{A}_{1}$ and $\mathcal{A}_{2}$ are $\mathcal{N}$ and $\mathcal{U},$
respectively, while the multipliers $\gamma _{1}$ and $\gamma _{2}$ are
identified with $-\mu /kT$ and $1/kT,$ respectively, where $T$ is the
temperature (in kelvin) and $\mu $ is the chemical potential. The function
(6.4) is then identified with the Massieu thermodynamic potential, and (6.3)
with the usual entropy of thermostatics since (6.5) implies the same equation

\begin{equation}
dS=\gamma _{1}d\mathcal{N}+\gamma _{2}d\mathcal{U}=\frac{1}{T}\left( d%
\mathcal{U}-\mu d\mathcal{N}\right)  \tag{6.6}
\end{equation}
as in macroscopic physics.

A similar formal structure, analogous to that of thermostatics, occurs for
any imperfectly known state at a given time, when the expectation values $%
\left\langle A_{i}\right\rangle =\mathcal{A}_{i}$ of some relevant set $%
R=\left\{ A_{i}\right\} $ of observables (as exemplified in \S 5) are the
only available information. A multiplier $\gamma _{i}$ is associated with
each given quantity $\mathcal{A}_{i}.$ The least biased state $D_{0}$ (eq.
6.1) is characterized equivalently by the set $\left\{ \mathcal{A}%
_{i}\right\} $ or by the set $\left\{ \gamma _{i}\right\} .$ The relations
(6.3) between these \textit{conjugate variables} $\left\{ \mathcal{A}%
_{i}\right\} $ and $\left\{ \gamma _{i}\right\} $ are generated from the
generalized thermodynamic potential $\Psi $ of eq.(6.4). The von Neumann
entropy (6.3) of the state (6.1), when regarded as a function of the
variables $\mathcal{A}_{i},$ appears as the Legendre transform of $\Psi
\left( \left\{ \gamma _{i}\right\} \right) .$ We denote it as

\begin{equation}
S_{R}\left( \left\{ \mathcal{A}_{i}\right\} \right) =S\left( D_{0}\right)
=\Psi +\underset{i}{\sum }\gamma _{i}\mathcal{A}_{i},  \tag{6.7}
\end{equation}
and it alternatively generates the relations between the conjugate sets $%
\left\{ \mathcal{A}_{i}\right\} $ and $\left\{ \gamma _{i}\right\} $ through

\begin{equation}
\frac{\partial S_{R}}{\partial \mathcal{A}_{i}}=\gamma _{i}.  \tag{6.8}
\end{equation}

We term $S_{R}$ the \textit{relevant entropy associated with the set }$%
R\equiv \left\{ A_{i}\right\} $ of relevant observables. By construction it
is the maximum value. $S\left( D_{0}\right) $ of the uncertainty --- as
measured by the von Neumann entropy $S\left( D\right) $ --- of the various
microscopic states $D$ which are equivalent as regards the expectation values

\begin{equation}
\left\langle A_{i}\right\rangle =\mathrm{Tr}\text{\thinspace }A_{i}D=\mathrm{%
Tr}\text{\thinspace }A_{i}D_{0}=\mathcal{A}_{i}  \tag{6.9}
\end{equation}
of the relevant observables, but which are inequivalent as regards the
irrelevant observables (those which lie outside $R).$ The relevant entropy
thus characterizes the \textit{randomness of an incomplete statistical
description }of the system, involving only the set $\mathcal{A}_{i}$ at the
considered time: it measures the amount of information which is missing when
only the quantities $\left\langle A_{i}\right\rangle =\mathcal{A}_{i}$ are
specified. While $D_{0}$ carries no other information than these expectation
values, the other density operators $D$ satisfying (6.9) carry an \textit{%
extra amount of information }which pertains to irrelevant observables and
which is measured by

\begin{equation}
S_{R}\left( \left\{ \mathcal{A}_{i}\right\} \right) -S\left( D\right) . 
\tag{6.10}
\end{equation}

The relevant entropy $S_{R}$ decreases as the set $R$ is enlarged. Consider
a statistical ensemble of systems all prepared in the same manner, by
controlling some set $R_{n}$ of observables. Imagine that we have no
information at the start and that we determine better and better the state
by learning the expectation values $\mathcal{A}_{i}$ for nested, larger and
larger sets $R_{1}\subset R_{2}\subset ...\subset R_{n}.$ (Or imagine that
conversely we begin with the full statistical knowledge of $R_{n}$ and
discard information by steps, until we keep only information on $R_{1}.)$
Each step corresponds to a positive gain (or loss) of information. This is
expressed by the inequalities

\begin{equation}
S_{R_{1}}>S_{R_{2}}>...>S_{R_{n}}  \tag{6.11}
\end{equation}
for the relevant entropies associated with each set.

An \textit{exercise }on relevant entropies can be found in ref. \cite
{balian1}, exerc. 3c.

\section{Reduction of the description}

The approach of \S 6 covers the foundations of thermostatics. New inference
problems arise, however, in non-equilibrium statistical mechanics.

We first have to assign a density operator $D\left( t_{0}\right) $ to the
system at the \textit{initial time } $t_{0}.$ Some macroscopic variables $%
\mathcal{A}_{i}$ are \textit{controlled externally }so as to prepare the
initial state, and this preparation is of course incomplete. We wish to
infer therefrom the complete statistical description which is required to
work out the microscopic approach of statistical physics. Such a question
has been solved in \S 6, where we showed that the initial state $D\left(
t_{0}\right) $ should have the form (6.1) in terms of the observables $A_{i}$
under experimental control. Actually these observables which can be handled
in practice to prepare the system may constitute only a \textit{subset} of
the chosen set $R$ of relevant observables. In spite of that, we note that $%
D\left( t_{0}\right) =D_{0}\left( t_{0}\right) $ is still expressed as (6.1)
in terms of the \textit{full set} $R,$ provided we associate a vanishing
multiplier $\gamma _{j}\left( t_{0}\right) =0$ with any relevant observable $%
A_{j}$ that is not controlled in the preparation. The values of the
multipliers $\gamma _{i}\left( t_{0}\right) $ associated with the controlled
data $\mathcal{A}_{i}\left( t_{0}\right) $ are still determined by (6.5),
while we infer from these controlled data the expectation values of the
remaining, non controlled relevant observables by means of

\begin{equation}
\left. A_{j}\left( t_{0}\right) =-\frac{\partial \Psi }{\partial \gamma _{j}}%
\right| _{\gamma _{j}=0}.  \tag{7.1}
\end{equation}

The next problem is to make reasonable \textit{statistical predictions }%
about any observable $B$ at any time $t$ subsequent to $t_{0}$ (or
symmetrically to make retrodictions for times $t$ earlier than $t_{0}).$
Starting from the initial condition $D\left( t_{0}\right) =D_{0}\left(
t_{0}\right) $ inferred above from the data $\mathcal{A}_{i}\left(
t_{0}\right) ,$ we determine $D\left( t\right) $ (at least in principle if
not in practice) by solving the Liouville--von Neumann equation (1.2). The
required quantities are then obtained as

\begin{equation}
\left\langle B\right\rangle _{t}=\mathrm{Tr}BD\left( t\right) .  \tag{7.2}
\end{equation}

The equation of motion (1.2) for $D\left( t\right) $ entails the constancy
of the von Neumann entropy (3.2), which keeps its initial value:

\begin{equation}
S\left[ D\left( t\right) \right] =S_{R}\left[ \left\{ \mathcal{A}_{i}\left(
t_{0}\right) \right\} \right] .  \tag{7.3}
\end{equation}
This equation expresses that the complete statistical description provided
by $D\left( t\right) $ \textit{conserves the information }that was initially
stored in the system through the controlled data $\mathcal{A}_{i}\left(
t_{0}\right) .$ This conservation is natural, owing to the reversibility of
the microscopic evolution. However, $D\left( t\right) $ has no reason to
keep the exponential form (6.1) that it had at the time $t_{0}.$ Hence, with
respect to the relevant set $R,D\left( t\right) $ is generally \textit{%
biased: }the information that it carries pertains not only to this set $R$
(as at the initial time $t=t_{0})$ but also to irrelevant observables. The
evolution thus generates some \textit{transfer of information from relevant
to irrelevant variables.}

Let us make this point clearer by \textit{getting rid of the irrelevant
information} at each time. We thus wish to study the relevant quantities.

\begin{equation}
\mathcal{A}_{i}\left( t\right) =\left\langle A_{i}\right\rangle _{t}=\mathrm{%
Tr}A_{i}D\left( t\right) ,  \tag{7.4}
\end{equation}
irrespective of the other variables. The method of \S 6 allows us to discard
the information carried by the latter, irrelevant variables. It provides us
with a state $D_{0}\left( t\right) ,$ which accounts for the values $%
\left\langle A_{i}\right\rangle _{t}$ given by (7.4) for the relevant
quantities and which, contrary to $D\left( t\right) ,$ is unbiased with
respect to the set $R.$ The values of $\left\langle B\right\rangle _{t}$
provided by $D_{0}\left( t\right) $ are those inferred from the knowledge of
the set $\mathcal{A}_{i}\left( t\right) $ only; they differ from the actual
values (7.2) which are inferred from the initial conditions $\mathcal{A}%
_{i}\left( t_{0}\right) $ and from the microscopic evolution. The missing
information associated with the sole knowledge of the set (7.4) is the
relevant entropy $S_{R}\left[ \left\langle \mathcal{A}_{i}\left( t\right)
\right\rangle \right] ,$ i.e., the von Neumann entropy $S\left[ D_{0}\left(
t\right) \right] $ of the unbiased state $D_{0}\left( t\right) .$

We are thus led to introduce, beside the complete statistical microscopic
description through $D\left( t\right) ,$ a reduced description in terms of a
simpler density operator $D_{0}\left( t\right) $ which carries the sole
information on the relevant set $R.$ At each time, we\textit{\ associate
with }$D\left( t\right) $ \textit{a reduced state }$D_{0}\left( t\right) $
which has the exponential form (6.1) and which satisfies the conditions

\begin{equation}
\mathrm{Tr}A_{i}D_{0}\left( t\right) =\mathrm{Tr}A_{i}D\left( t\right) =%
\mathcal{A}_{i}\left( t\right) .  \tag{7.5}
\end{equation}
The reduced state $D_{0}\left( t\right) $ accompanies $D\left( t\right) $ in
its motion. The parameters $\gamma _{i}\left( t\right) $ of (6.1) which
determine $D_{0}\left( t\right) $ are obtained from (7.5). The
correspondence between the sets $\left\{ \gamma _{i}\left( t\right) \right\} 
$ and $\left\{ \mathcal{A}_{i}\left( t\right) \right\} $ is implemented by
(6.5) or (6.8), so that we can characterize the reduced state equivalently
by the set $\left\{ \mathcal{A}_{i}\left( t\right) \right\} ,$ by the set $%
\left\{ \gamma _{i}\left( t\right) \right\} $ or by the density operator $%
D_{0}\left( t\right) .$

The information carried by this reduced description is smaller than that
carried by the full description through $D\left( t\right) .$ The difference

\begin{equation}
S\left[ D_{0}\left( t\right) \right] -S\left[ D\left( t\right) \right] =S_{R}%
\left[ \left\{ \mathcal{A}_{i}\left( t\right) \right\} \right] -S_{R}\left[
\left\{ \mathcal{A}_{i}\left( t_{0}\right) \right\} \right] ,  \tag{7.6}
\end{equation}
where we used (7.3), \textit{measures the irreversibility }with respect to
the set $R.$ It is the amount of \textit{information which has leaked }from
the relevant to the irrelevant set. It is also interpreted as the \textit{%
transfer of disorder }from the irrelevant to the relevant set. According to
(7.3), the total disorder $S\left[ D\left( t\right) \right] $ remains
constant, while (7.6) shows that the disorder associated with the relevant
set has increased between the times $t_{0}$ and $t.$

Altogether, we have been able to relate the macroscopic description in terms
of the sole variables $\mathcal{A}_{i}\left( t\right) $ to the microscopic,
detailed description of statistical mechanics. The following scheme was
brought out:

\begin{equation}
\left\{ \mathcal{A}_{i}\left( t_{0}\right) \right\} \Rightarrow D\left(
t_{0}\right) \leftrightarrow D\left( t\right) \mapsto \left\{ \mathcal{A}%
_{i}\left( t\right) \right\} \Rightarrow D_{0}\left( t\right) .  \tag{7.7}
\end{equation}
From the \textit{data} $\left\{ \mathcal{A}_{i}\left( t_{0}\right) \right\} $
which are controlled at the initial time, we first \textit{inferred the state%
} \textit{\ }$D\left( t_{0}\right) =D_{0}\left( t_{0}\right) $ at that time
by means of the maximum entropy criterion. The solution of the \textit{%
Liouville--von Neumann equation}, which is the basic dynamical equation in
statistical mechanics, related $D\left( t\right) $ to $D\left( t_{0}\right) $
reversibly. We then found the \textit{relevant quantities} $\left\{ \mathcal{%
A}_{i}\left( t\right) \right\} $ through (7.4). Eventually we \textit{%
inferred }$D_{0}\left( t\right) $ from the latter set, using again the
maximum entropy criterion. This last step will be used in \S 8 to eliminate
explicitly the irrelevant variables. The introduction of $D_{0}\left(
t\right) $ has also been necessary to define the relevant entropy, which
according to (7.6) characterizes the loss of information in the step $%
D\left( t\right) \mapsto \left\{ \mathcal{A}_{i}\left( t\right) \right\} $
or the \textit{irreversibility of the correspondence from }$\left\{ \mathcal{%
A}_{i}\left( t_{0}\right) \right\} $ \textit{to} $\left\{ \mathcal{A}%
_{i}\left( t\right) \right\} .$

\section{The projection method}

Macroscopic descriptions of dynamical processes rely on more or less
phenomenological equations for the macroscopic variables $\left\{ \mathcal{A}%
_{i}\left( t\right) \right\} .$ We now consider a major problem of
statistical physics, namely justifying these macroscopic equations from the
microscopic, more fundamental theory. To this is aim, we have to get rid of
the irrelevant microscopic variables so as to work out the scheme (7.7) and
to find equations of motions for the set $\left\{ \mathcal{A}_{i}\left(
t\right) \right\} $ only. The projection method of Nakajima and Zwanzig
(4.5) achieves this goal, at least formally but quite generally. We sketch
it here in order to shed light on the meaning of varied macroscopic
dynamical equations.

We first note that the knowledge of the time-dependence of the set $\left\{ 
\mathcal{A}_{i}\left( t\right) \right\} $ is equivalent to that of the
reduced state $D_{0}\left( t\right) .$ Indeed, we can deduce $\mathcal{A}%
_{i} $ from $D_{0}$ through (7.5). Conversely $D_{0},$ as expressed by
(6.1), (6.3), depends on time through the variables $\left\{ \gamma
_{i}\right\} ,$ which in turn are related to the variables $\left\{ \mathcal{%
A}_{i}\right\} $ by (6.5) or (6.8). We shall in the following regard $D_{0}$
as a function of the set $\left\{ \mathcal{A}_{i}\right\} ,$ and as a
function of time through this set.

The time-dependence of the reduced state $D_{0}$ that we thus seek for
should be obtained from the Liouville--von Neumann equation of motion for
the full $D.$ We shall work out this derivation in a Liouville
representation (\S 2), where both the density operators and the observables
are regarded as vectors in two conjugate vector spaces. In such a
representation the latter equation of motion as well as the correspondence
from $D$ to $D_{0}$ will be expressed by formally simple equations. On the
one hand the Liouville--von Neumann equation is written as (2.3) in terms of
the liouvillian superoperator $\mathsf{L}.$

On the other hand, $D_{0},$ which is a function of the set $\left\{ \mathcal{%
A}_{i}\right\} $ through (6.1), (6.3) and (6.8), follows from $D$ through
the evaluation of the set $\left\{ \mathcal{A}_{i}\right\} $ by means of
(7.4). It can be shown that this construction amounts, in a Liouville
representation where $D$ and $D_{0}$ are regarded as vectors in the
Liouville space, to a \textit{projection of} $D$ \textit{onto the surface}
(6.1):

\begin{equation}
D_{0}=\mathsf{P}D.  \tag{8.1}
\end{equation}
The \textit{maximum entropy }criterion is thus implemented as a \textit{%
projection.} In a Hilbert space representation, the projection superoperator 
$\mathsf{P}$ is a tensor with $2\times 2$ indices, which acts in the same
way (2.4) as the liouvillian on the density matrix $D.$ The projection $%
\mathsf{P}$ depends on the variables $\left\{ \mathcal{A}_{i}\right\} .$ It
acts on its left hand side upon the observables, and satisfies in particular

\begin{equation}
A_{i}\mathsf{P}=A_{i}\,,\,\,I\mathsf{P}=I.  \tag{8.2}
\end{equation}
The projection thus keeps the \textit{relevant set }$R$ \textit{of
observables invariant. }These facts are easily checked from the expression
of $\mathsf{P},$

\begin{equation}
\mathsf{P}=D_{0}\otimes I+\underset{i}{\sum }\frac{\partial D_{0}}{\partial 
\mathcal{A}_{i}}\otimes \left( A_{i}-\mathcal{A}_{i}I\right) ,  \tag{8.3}
\end{equation}
which however we shall not need below. Eq. (8.3) exhibits $\mathsf{P}$ as a
sum of elementary projections, each of which is the tensor product of a
state-like vector and an observable-like vector, that lie in the conjugate
Liouville spaces; the partial derivative $\partial D_{0}/\partial \mathcal{A}%
_{i}$ is evaluated by regarding $D_{0}$ as a function of the variables $%
\left\{ \mathcal{A}_{i}\right\} .$

The \textit{complementary projection }superoperator $\mathsf{Q}$ onto the
irrelevant space is defined as

\begin{equation}
\mathsf{Q}=\mathsf{I}-\mathsf{P},  \tag{8.4}
\end{equation}
where $\mathsf{I}$ is the unit superoperator in the Liouville space. The
properties $\mathsf{P}^{2}=\mathsf{P},\mathsf{Q}^{2}=\mathsf{Q}$ and $%
\mathsf{PQ\,}=0$ show that the Liouville formalism helps separating the
relevant and the irrelevant components. The density operator $D$ can thereby
be split at each time into its\textit{\ reduced relevant part }$D_{0}$ and
its \textit{irrelevant part }$D_{1}$ defined by

\begin{equation}
D_{1}=\mathsf{Q}D.  \tag{8.5}
\end{equation}

We wish to find the time-dependence of $D_{0}$ from the Liouville--von
Neumann equation (2.3) which deals with the full $D.$ We take therefore the
time-derivative of (8.1), using (2.3). This yields

\begin{equation}
\frac{dD_{0}}{dt}=\frac{d\mathsf{P}}{dt}D_{0}+\mathsf{PL}D_{0}+\mathsf{PL}%
D_{1},  \tag{8.6}
\end{equation}
where the time-dependence of $\mathsf{P}$ occurs through the variables $%
\left\{ \mathcal{A}_{i}\right\} .$ Apart from its last term where $D_{1}$ is
as get unknown, (8.6) is a differential equation for $D_{0}$ or for the set $%
\left\{ \mathcal{A}_{i}\right\} .$ It can be written equivalently, using
(7.5), as the set 
\begin{equation}
\frac{d\mathcal{A}_{i}}{dt}=\text{\textbf{(}}A_{i}\,\text{\textbf{%
;\thinspace }}\mathsf{L}D_{0}\text{\textbf{)}}\mathbf{\,}+\,\text{\textbf{(}}%
A_{i}\,\text{\textbf{;\thinspace }}\mathsf{L}D_{1}\text{\textbf{)}}, 
\tag{8.7}
\end{equation}
where we made use of the notation (2.2) for a scalar product in the
Liouville space and where we took (8.2) into account.

In order to \textit{eliminate }$D_{1}$ from (8.7), we write the equation of
motion for $D_{1}$, by taking the time-derivation of (8.5), which yields

\begin{equation}
\frac{dD_{1}}{dt}+\frac{d\mathsf{P}}{dt}D_{1}-\mathsf{QL}D_{1}=\mathsf{QL}%
D_{0}.  \tag{8.8}
\end{equation}
The equations (8.6) and (8.8) couple the projected parts $D_{0}$ and $D_{1}$
of $D$. As (8.8) is linear in $D_{1},$ with coefficients depending on $%
D_{0}, $ it can be solved formally by introducing its Green's function 
\textsf{W}$\left( t,t^{\prime }\right) .$ This quantity is a superoperator,
which is the solution of

\begin{equation}
\left( \frac{\partial }{\partial t}+\frac{d\mathsf{P}}{dt}-\mathsf{QL}%
\right) \mathsf{W}\left( t,t^{\prime }\right) =\mathsf{Q}\delta \left(
t-t^{\prime }\right) ,  \tag{8.9}
\end{equation}
with $\mathsf{W}\left( t,t^{\prime }\right) =0$ for $t<t^{\prime }.$ Using
the initial condition $D_{1}(t_{0})=0$ that results from $D$ $%
(t_{0})=D_{0}(t_{0})$ (\S 7), we thus express $D_{1}$ in terms of $D_{0}$ as

\begin{equation}
D_{1}\left( t\right) =\int_{t_{0}}^{t}dt^{\prime }\,\mathsf{W}\left(
t,t^{\prime }\right) \mathsf{L}D_{0}\left( t^{\prime }\right) .  \tag{8.10}
\end{equation}

Insertion of (8.10) into (8.7) achieves one goal, finding a set of dynamical
equations for the variables $\mathcal{A}_{i}\left( t\right) $ alone:

\begin{equation}
\frac{d\mathcal{A}_{i}}{dt}=\text{\textbf{(}}\mathcal{A}_{i}\,\text{\textbf{%
;\thinspace }}\mathsf{L}D_{0}\left( t\right) \text{\textbf{)}}+\text{\textbf{%
(}}\mathcal{A}_{i}\,\text{\textbf{;\thinspace }}\int_{t_{0}}^{t}dt^{\prime
}\,\mathsf{LW}\left( t,t^{\prime }\right) \mathsf{L}D_{0}\left( t^{\prime
}\right) \text{\textbf{)},}  \tag{8.11}
\end{equation}
where $\mathsf{W}$ was defined by (8.9). The elimination of the irrelevant
variables has resulted in a set of \textit{integro-differential equations}
which are \textit{non-linear} and \textit{non-markovian.} Indeed the last of
(8.11) does not depend on $D_{0}\left( t\right) $ alone as does the first
term, but on $D_{0}$ \textit{at earlier times,} directly through $%
D_{0}\left( t^{\prime }\right) $ and indirectly through the projections $%
\mathsf{P}$ and $\mathsf{Q}$ which enter the definition (8.9) of $\mathsf{W.}
$ The evolution of the set $\left\{ \mathcal{A}_{i}\right\} $ thus involves
a \textit{memory, }which is characterized by the \textit{memory kernel }$%
\mathsf{W}\left( t,t^{\prime }\right) .$ The latter superoperator \textit{%
depends on the history} of $\mathcal{A}_{i}$ between the times $t^{\prime }$
and $t.$

This first term in (8.11) represents the \textit{direct coupling }of the
observables within the relevant set $R;$ it involves the components $\mathsf{%
PLP}$ of the liouvillian. The last, retarded term is interpreted as the
effect of an \textit{indirect coupling through the irrelevant observables.}
Since $\mathsf{W}$ satisfies

\begin{equation}
\mathsf{W}\left( t,t^{\prime }\right) =\mathsf{Q}\left( t\right) \mathsf{W}%
\left( t,t^{\prime }\right) \mathsf{Q}\left( t^{\prime }\right) ,  \tag{8.12}
\end{equation}
and since its definition (8.9) involves only the components $\mathsf{QLQ}$
of the liouvillian, it describes the \textit{time-evolution of the
irrelevant set }of variables that we have eliminated. This irrelevant set is
coupled in (8.11) to the set $R$ of interest, at the time $t$ through the
cross components $\mathsf{PLQ}$ of the liouvillian, at the earlier times $%
t^{\prime }$ through its components $\mathsf{QLP}$.

We noted that the von Neumann entropy $S\left( D\right) ,$ associated
through (3.2) with the complete statistical description $D,$ remains
constant during the evolution generated by (1.2), expressing that this
evolution conserves information or disorder. We have here to \textit{%
associate with the reduced equation} of motion (8.11) \textit{the relevant
entropy} $S_{R}\left( \left\{ \mathcal{A}_{i}\right\} \right) $ expressed by
(6.7), which characterizes the disorder in the relevant set $R.$ Its time
variation, obtained from (6.8) and (8.11) as

\begin{eqnarray}
\frac{dS_{R}}{dt} &=&\underset{i}{\sum }\gamma _{i}\frac{d\mathcal{A}_{i}}{%
dt}  \notag \\
&=&\text{\textbf{(}}-\ln D_{0}\left( t\right) \text{\textbf{;}}%
\int_{t_{0}}^{t}dt^{\prime }\,\mathsf{LW}\left( t,t^{\prime }\right) \mathsf{%
L}D_{0}\left( t^{\prime }\right) \text{\textbf{)}},  {(8.13)}
\end{eqnarray}
has no reason to vanish. In fact, the positivity of (7.6) shows that, at
least between the times $t_{0}$ and $t,$ the relevant entropy increases. The
expression (8.13), after integration over time, relates this loss of
information to the coupling of the relevant set $R$ with the irrelevant
observables.

\section{Short-memory approximation}

Our whole construction relied on the choice of the relevant set $R,$ which
guided by macroscopic physics, itself based on experiment and on
phenomenological theory. The macroscopic dynamical equations for the
variables $\mathcal{A}_{i}\left( t\right) $ usually have the form of
differential equations. In contrast, the exact equations (8.11) have a
different structure, involving memory effects for whatever choice of $R.$ A
further step is thus needed to found the macroscopic equations upon the
microscopic theory.

We wish the last term of (8.11) to depend approximately only on the set $%
\left\{ \mathcal{A}_{i}\left( t\right) \right\} $ at the considered time $t,$
not at earlier times. This can be achieved if the choice of the relevant set 
$R$ satisfies some conditions about \textit{time-scales.} Actually, the main
contribution to the integral over $t^{\prime }$ in (8.11) should arise from
times $t^{\prime }$ sufficiently close to the upper bound $t$ so that $%
D_{0}\left( t\right) $ \textit{does not differ significantly from }$%
D_{0}\left( t\right) .$ This requires the weight

\begin{equation*}
\mathsf{P}\left( t\right) \mathsf{L\,W}\left( t,t^{\prime }\right) \mathsf{%
L\,P}\left( t^{\prime }\right) ,
\end{equation*}
which multiplies $D_{0}\left( t^{\prime }\right) $ and which accounts for
the dynamics of the irrelevant variables, to \textit{decrease sufficiently
fast }as $t-t^{\prime }$ increases. The relevant set should thus have been
chosen by \textit{selecting the slowest variables.} We understand the
possibility of a short time-range $\tau $ for the expression (9.1) by noting
that it is equal to $\mathsf{P}\left( t\right) \mathsf{LQ}\left( t\right) 
\mathsf{LP}\left( t\right) $ when $t=t^{\prime },$ and that for $t>t^{\prime
}$ it includes a sum of a large number of terms, associated with the
evolution of the irrelevant degrees of freedom which oscillate rapidly and
interfere destructively as $t-t^{\prime }$ increases. If (9.1) is thus
negligible beyond the delay $\tau $, the \textit{memory }\ in the evolution
of the relevant set due to their coupling with the irrelevant ones is 
\textit{lost} after the time-lapse $\tau .$

We can then replace $D_{0}\left( t^{\prime }\right) $ by $D_{0}\left(
t\right) $ in the equations (8.11), which become approximately

\begin{equation}
\frac{d\mathcal{A}_{i}}{dt}\simeq \text{\textbf{(}}A_{i}\,\text{\textbf{%
;\thinspace }}\left[ \mathsf{L}+\mathsf{LK}\left( t\right) \mathsf{L}\right]
D_{0}\left( t\right) \text{\textbf{)}}\mathbf{;}  \tag{9.2}
\end{equation}
we defined $\mathsf{K}\left( t\right) $ as

\begin{equation}
\mathsf{K}\left( t\right) =\int_{t_{0}}^{t}dt^{\prime }\,\mathsf{W}\left(
t,t^{\prime }\right) .  \tag{9.3}
\end{equation}
The lower bound $t_{0}$ in (9.2) is not significant since the integrand is
negligible outside the interval $t>t^{\prime }\gtrsim t-\tau .$ The
superoperator $\mathsf{K}$ acts only on the irrelevant space since $\mathsf{K%
}\left( t\right) =\mathsf{Q}\left( t\right) \mathsf{K}\left( t\right) 
\mathsf{Q}\left( t\right) ,$ as seen from (8.12) and from the slow variation
of $\mathsf{Q}\left( t\right) $ on the scale $\tau .$ The bracket in (9.2)
can equivalently be replaced by $\mathsf{P}\left[ \mathsf{L+LKL}\right] 
\mathsf{P},$ which shows that it acts on the relevant space only.

The projection method, completed by this short-memory approximation, has
thus led to the equations of motion (9.2) which provide a \textit{%
microscopic basis for the macroscopic dynamics.} The possibility of
describing the system by differential equations for some partial set of
relevant variables, without considering the remaining irrelevant microscopic
variables, relied on the recognition of different characteristic time-scales
for the two sets. The influence on the macroscopic dynamics of the
irrelevant variables which have been eliminated then occurs through the
factor $\mathsf{PLKLP}$ in (9.2). The method is general and flexible, as
examples will show. It is however formal, since the solution of the equation
(8.9) for $\mathsf{W}$ is at least as difficult as that of the
Liouville--von Neumann equation. Further approximation (such as perturbative
expansions) specific to the considered problem, are therefore needed to
evaluate $\mathsf{K}$ from first principles. A rough approximation consists
in estimating only the memory-time $\tau $ and in taking $\mathsf{K}\simeq
\tau \mathsf{Q}.$

When the approximation (9.2) is valid, the \textit{dissipation }(8.13) has
the instantaneous form

\begin{equation}
\frac{dS_{R}}{dt}=\text{\textbf{(}}-\ln D_{0}\,\text{\textbf{;\thinspace }}%
\mathsf{LKL}D_{0}\text{\textbf{)}}\mathbf{.}  \tag{9.4}
\end{equation}
It is \textit{positive at each time,} expressing a \textit{continuous flow
of disorder} from the irrelevant to the relevant variables. Although
dissipation appears from (8.13) to be a memory effect, it is in fact a 
\textit{short-memory effect,} taking place on the time-scale $\tau .$ The
reconciliation of \textit{microscopic reversibility} with \textit{%
macroscopic irreversibility }thus requires the existence of two different
time-scales.

\section{The thermodynamic entropy}

We have seen in \S 5 that the macroscopic state of a system evolving in a
thermodynamic regime is characterized by a set of \textit{local variables} $%
\mathcal{A}_{ia}$ associated, for each subsystem located at $a$ with some or
other quantity $i.$ These quantities may flow from a subsystem \textit{a} to
another, and thermodynamics deals with such a \textit{transport} \cite
{callen}: thermal conduction corresponds to flow of energy, diffusion or
electric current to flow of particles, fluid dynamics to flow of momentum.
The thermodynamic equations provide the time-dependence of the variables $%
\mathcal{A}_{ia}.$ They can be decomposed into: (i) \textit{Conservation laws%
} relating the time-derivatives $d\mathcal{A}_{ia}/dt$ to \textit{fluxes }%
from each subsystem to its neighbours. (ii) For each subsystem \textit{a},
relations between the variables $\mathcal{A}_{ia}$ and their conjugate 
\textit{local intensive variables }$\gamma _{ja},$ such as temperature,
chemical potential or velocity; these relations are at each point and each
time the same as in thermostatics. (iii) Definition of \textit{affinities}
as differences $\gamma _{ia}-\gamma _{ib}$ or as gradients of the intensive
variables. $\left( iv\right) $ \textit{Responses }of the fluxes to the
affinities.

The equations for the set $\left\{ \mathcal{A}_{ia}\right\} $ characterize
the exchanges which occur between the subsystems. Such exchanges take place
on a time-scale which is large compared the time-scale $\tau $ of the
microscopic processes. In the present case, the latter time $\tau $
corresponds to the delay after which the effect of the perturbation of a
microscopic variable has been forgotten. It is thus the \textit{relaxation
time }which brings the system to \textit{local equilibrium}. Thereafter the
evolution slows down, and proceeds in the thermodynamic or hydrodynamic
regime where the dynamics involves the local quantities $\mathcal{A}_{ia}$
only. In a gas, $\tau $ is of the order of the delay between two successive
collisions of a particle.

The occurence of a larger time-scale for the thermodynamic variables $%
\mathcal{A}_{ia}$ is related to the fact that these variables are \textit{%
conserved} or nearly conserved. Indeed, the conservation laws imply that $%
\mathcal{A}_{ia}$ can increase only if $\mathcal{A}_{ib}$ in a neighbouring
subsystem $b$ decreases, under the effect of a flux from $b$ to $a.$ This
requires a coupling between the subsystems $a$ and $b.$ However, due to the
macroscopic size of the subsystems, such a coupling is an interface effect
which may be relatively weak; this hinders the fluxes and hence the
time-variation of the variables $\mathcal{A}_{ia}.$

The microscopic incomplete description associated with the set of
thermodynamic variables $\mathcal{A}_{ia}$ is based on a reduced density
operator $D_{0}$ of the form (6.1). Separation of the observables $\mathcal{A%
}_{ia}$ according to the subsystem \textit{a} that they refer to expresses $%
D_{0}$ as a tensor product

\begin{equation}
D_{0}=\underset{a}{\prod }D_{a},\,D_{a}=\exp \left[ -\Psi _{a}-\underset{i}{%
\sum }\gamma _{ia}A_{ia}\right]  \tag{10.1}
\end{equation}
of elementary density operators $D_{a},$ each of which refers to one of the
subsystems. In fact $D_{a}$ has exactly the \textit{same form as }the
density operator of the system \textit{a} alone \textit{in thermostatic
equilibrium.} For each subsystem the relations between the variables $%
\mathcal{A}_{ia}$ and $\gamma _{ja}$ are thus the same as in thermostatics.
We can therefore identify the multipliers $\gamma _{ia}$ of (10.1) with the
corresponding intensive variables of thermodynamics. We can also identify
the set (9.2) with the thermodynamical equations of motion.

Moreover the factorization (10.1) of $D_{0}$ implies that the relevant
entropy (6.7) associated with the thermodynamic variables is simply the sum

\begin{equation}
S_{R}\left( \left\{ \mathcal{A}_{ia}\right\} \right) =\underset{a}{\sum \,}%
S\left( D_{a}\right)  \tag{10.2}
\end{equation}
of the von Neumann entropies $S\left( D_{a}\right) ,$ each of which is equal
to the thermostatic entropy of the subsystem $a.$ This allows us to identify
the \textit{relevant entropy }$S_{R}\left( \left\{ \mathcal{A}_{ia}\right\}
\right) $ relative to the thermodynamic variables with the \textit{entropy }$%
S_{\text{\textrm{th}}}$\textit{\ of thermodynamics.}

The \textit{Second Law }of thermostatics refers to a system in local
equilibrium at the initial time $t_{0},$ in global equilibrium at a final
time $t_{1}.$ In the interval, the evolution does not need to take place in
the thermodynamic regime. We may for instance deal with an explosive
chemical reaction, or a shock wave, and the short-memory approximation of \S
9 is thus not necessarily valid. However, while the von Neumann entropy $S%
\left[ D\left( t\right) \right] $ remains constant, the positivity of (7.6)
implies that $S_{R}\left[ \left\{ \mathcal{A}_{ia}\left( t_{1}\right)
\right\} \right] >S_{R}\left[ \left\{ \mathcal{A}_{ia}\left( t_{0}\right)
\right\} \right] ,$ which proves microscopically the Second Law.

If the evolution takes place in the thermodynamic regime according to (9.2),
the dissipation (9.4) is at each time positive, which means that $S_{\mathrm{%
th}}$ keeps increasing. This property is the dynamical form of the Second
Law, expressed by the \textit{Clausius--Duhem inequality}

\begin{equation}
\frac{dS_{\mathrm{th}}}{dt}\geqslant 0.  \tag{10.3}
\end{equation}

\section{The Boltzmann entropy}

We turn to the Boltzmann description of a gas (\S 5). Here the relevant
observables $A_{i}$ are the \textit{single-particle random variables;} their
expectation values are the set $\left\{ \mathcal{A}_{i}\right\} =f\left( 
\mathbf{r},\mathbf{p}\right) ,$ i.e., the density of particles in the
single-particle phase space. This description is more detailed than the
thermodynamic description of the gas, which would rely only on the densities
of particles, of energy and of momentum in the ordinary space $\mathbf{r.}$
It applies to more general regimes than the thermodynamic regime. In
particular, in the \textit{ballistic regime,} the time between collisions
which characterizes the thermalization is larger than the time during which $%
f\left( \mathbf{r},\mathbf{p}\right) $ changes significantly; the mean free
path is larger than the distance over which $f$ changes. No local
temperature, no hydrodynamic flow can then be defined.

While $f\left( \mathbf{r},\mathbf{p}\right) $ plays the r\^{o}le of the
macroscopic variables $\left\{ \mathcal{A}_{i}\right\} ,$ the \textit{%
irrelevant variables are here }the \textit{correlations} between the
particles of the gas. Such correlations are created by collisions; the
characteristic time associated with their dynamics is the duration of a
collision. Over time-scales larger than this very brief duration, and over
distances larger than the range of the interactions, the dynamics of $%
f\left( \mathbf{r},\mathbf{p;}t\right) $ is governed by the
semi-phenomenological \textit{Boltzmann equation}

\begin{equation}
\frac{\partial f\left( \mathbf{r},\mathbf{p;\,}t\right) }{\partial t}=-\frac{%
\mathbf{p}}{m}\cdot \nabla _{r}\,f+\mathcal{I\,}\left[ f\right] .  \tag{11.1}
\end{equation}
This equation of motion for a gas is adequate under conditions much more
general than the equations of thermodynamics. Like the latter ones it should
be supplemented by boundary conditions accounting for the vessel in which
the gas is enclosed. We shall consider it as a macroscopic equation and $%
f\left( \mathbf{r},\mathbf{p}\right) $ as a continuous set of macroscopic
variables, whereas the Liouville equation for the full $D$ is the exact
microscopic equation. The last term $\mathcal{I\,}\left[ f\right] $ is the
collision integral, a quadratic functional of $f$ involving integration over
momenta.

By introducing the quantity

\begin{equation}
H\left( t\right) \equiv \int d^{3}\mathbf{r\,}d^{3}\mathbf{p\,}f\,\ln \,f, 
\tag{11.2}
\end{equation}
Boltzmann proved that his equation (11.1) satisfies the \textit{H-theorem, }%
namely

\begin{equation}
\frac{dH}{dt}\leqslant 0.  \tag{11.3}
\end{equation}
This exhibits the irreversibility of the Boltzmann equation, which contrasts
with the reversibility of the more basic Liouville equation.

At the microscopic level the Boltzmann description relies on a reduced
density $D_{0}$ in the 6$N$-dimensional phase space, which is the classical
equivalent of (6.1). The random variables $A_{i}$ entering it are the
classical single-particle quantities, so that $D_{0}$ describes an \textit{%
uncorrelated state.} The projection (8.1) amounts here to \textit{chop off
the correlations }from $D,$ while keeping the expectation values $f\left( 
\mathbf{r},\mathbf{p}\right) $ of single-particle quantities invariant. The
Boltzmann equation is then identified with the result (9.2) of the
projection method, the short-memory approximation being here justified by
the short duration of the collisions. The two terms of (11.1) actually
correspond to the two terms of (9.2).

The relevant entropy associated with the Boltzmann variables $f\left( 
\mathbf{r},\mathbf{p}\right) $ is found by taking the classical limit of
(6.4), (6.5), (6.7), where the trace is replaced by the measure (2.6). We
obtain

\begin{equation}
S_{\mathrm{B}}\equiv S_{R}\left( f\right) =\int d^{3}\mathbf{r\,}d^{3}%
\mathbf{p}\,f\left( \mathbf{r},\mathbf{p}\right) \left[ 1-\ln h^{3}f\left( 
\mathbf{r},\mathbf{p}\right) \right] .  \tag{11.4}
\end{equation}
This expression defines the \textit{Boltzmann entropy }$S_{\mathrm{B}},$
which is relevant for the dynamics of a gas in whatever regime, provided the
range of the interparticle potential is short compared to the distances
between particles. Since the integral of $f$ is the particle number, the
Boltzmann entropy $S_{\mathrm{B}}$ is proportional to Boltzmann's $H,$
within the sign and within an additive constant. Thus the $H$-theorem
expresses merely the increase (9.4) of the Boltzmann entropy.

The $H$-theorem should not be confused with the Second Law. On the one hand,
it applies only to \textit{gases.} On the other hand, it covers much \textit{%
more general situations, }since in thermostatics the Second Law requires
local equilibrium for the initial state, global equilibrium for the final
state, and since in thermodynamics the Clausius--Duhem inequality requires
local equilibrium at all times.

The \textit{thermodynamics of gases} is recovered from the Boltzmann
description in regimes for which $f$ varies slowly in space on the scale of
the mean free path, slowly in time on the scale of the delay between
collisions. In this limit, it can be shown that of $f\left( \mathbf{r},%
\mathbf{p,}t\right) $ does not differ much from a single-particle density in
phase space having in terms of $\mathbf{p}$ the \textit{maxwellian form}

\begin{equation}
f_{0}\left( \mathbf{r},\mathbf{p,}t\right) =\exp \left[ -\gamma _{1}-\gamma
_{2}\frac{p^{2}}{2m}-\mathbf{\gamma }_{3}\cdot \mathbf{p}\right] , 
\tag{11.5}
\end{equation}
\textit{\ }where $\gamma _{1},\gamma _{2}$ and $\mathbf{\gamma }_{3}$ depend
on $\mathbf{r}$ and $t.$ The latter quantities are interpreted as the local
intensive variables in the local equilibrium regime of thermodynamics, for
instance $1/k\gamma _{2}$ as the local temperature (the energy of the gas
includes practically the kinetic energy only of the particles). The
Boltzmann entropy (11.4) reduces to the thermodynamic entropy $S_{\mathrm{th}%
},$ the $H$-theorem to the thermodynamic dissipation (10.3). The
thermodynamic equations of motion can also be obtained from the Boltzmann
equation (11.1). The \textit{Chapman--Enskog method,} on which this
derivation is based, appears as an application of the \textit{projection
method. }Here $f$ is regarded as the set of microscopic variables. We thus
replace in the general formalism $D$ by $f,$ and the Liouville--von Neumann
equation by the Boltzmann equation (see for instance \cite{balian1}, chap.
15). Like $D$ in eqs. (8.1) and (8.5), $f$ is split into $f=f_{0}+f_{1},$
where $f_{0},$ of the form (11.5), is obtained from $f$ by means of a
projection in the space of the functions of $\mathbf{p}$. The elimination of 
$f_{1}$ is made here feasible by regarding $f_{1}$ as small compared to $%
f_{0}$ in the considered regime, and $\mathrm{it}$ results in the
Navier--Stokes and Fourier equations.

\section{Conclusion: the multiplicity of entropies}

The above example of a gas exhibits the occurrence in non-equilibrium
statistical mechanics of several different entropies for a single system.

(i) The \textit{von Neumann entropy }$S\left( D\right) $ defined by (3.2) is
associated with the complete statistical description of the gas. It accounts
in particular for the order carried by the correlations between its
particles, which develop as collisions take place, even if they are absent
at the initial time. This entropy remains constant, expressing the
reversibility of the microscopic evolution and the conservation of the
initial information.

(ii) The \textit{Boltzmann entropy} $S_{\mathrm{B}}$ defined by (11.4) is
the relevant entropy associated with the single-particle properties only, in
an incomplete description disregarding all the correlations. Its increase in
time, the $H$-theorem, expresses the irreversible loss of information, or of
order, which take place from the single-particle quantities towards the
correlations. Actually, although the interactions an essential in the
dynamics of the gas, the correlations which may exist between a particle and
other ones are not effective when this particle undergoes a new collision;
this is Boltzmann's Stosszahlansatz, that he used to justify his equation
(11.1).

(iii) The \textit{thermodynamic entropy} $S_{\mathrm{th}}$ defined by (10.2)
is the relevant entropy associated with only the locally conserved densities
of particles, of energy and of momentum at each point. In the thermodynamic
regime where $f$ has the form (11.5), the short-memory approximation is
valid and it ensures that $S_{\mathrm{th}}$ continuously increases. This
expresses the dissipation due to viscosity and to thermal conduction.

(iv) The \textit{thermostatic entropy} $S_{\text{\textrm{eq}}}\left( 
\mathcal{N},\mathcal{U}\right) $ defined in \S 6 can also be regarded as the
relevant entropy associated with the total particle number and the total
energy. These quantities are the globally conserved ones; momentum is
locally conserved but not globally due to the collisions onto the walls, it
vanishes at equilibrium. This definition holds at any time for a gas off
equilibrium, with $\mathcal{N}$ and $\mathcal{U}$ keeping their initial
values. However $S_{\text{\textrm{eq}}},$ which does not vary, has no
physical interest, except after the very long, global relaxation time has
elapsed. The gas has then been brought to thermostatic equilibrium through
the evolution generated by the Boltzmann equation.

Because the above four descriptions are more and more incomplete, their
three relevant entropies satisfy the inequalitites (6.11), to wit,

\begin{equation}
S_{\text{\textrm{eq}}}>\,S_{\mathrm{th}}>S_{\mathrm{B}}>\,S\left( D\right) .
\tag{12.1}
\end{equation}
The entropies $S_{\mathrm{th}}$ and $S_{\mathrm{B}}$ thus vary between the
constant lower and upper bounds $S\left( D\right) $ and $S_{\text{\textrm{eq}%
}},$ and $S_{\mathrm{B}}$ never decreases. After local equilibrium is
established, $S_{\mathrm{B}}$ and $S_{\mathrm{th}}$ are nearly equal and
both tend to $S_{\text{\textrm{eq}}}$ when the system reaches global
equilibrium; as we just saw, $S_{\text{\textrm{eq}}}$ is of interest only
afterwards. Likewise, $S_{\mathrm{th}}$ has no interest before local
equilibrium is reached, that is, during some time-lapse of the order of the
delay between collisions. In this initial fast regime, the thermodynamic
variables are inadequate. They obey only an impracticable equation of the
type (8.11) including retardation effects, and not yet a thermodynamic
equation of the type (9.2). Accordingly, nothing forces $dS_{\mathrm{th}}/dt$
to be always positive at this stage of the evolution.

The contrast between the constancy of $S\left( D\right) $ and the increase
of $S_{\mathrm{B}}$ is better understood by introducing still other relevant
entropies \cite{mayer}. Starting from the Boltzmann description based on
single-particle quantities only, a hierarchy of \textit{more and more
detailed descriptions }are introduced by including two-particle
correlations, then three-particle correlations, and so on. The sequence of
associated relevant entropies $S_{2},S_{3},...$ satisfies the inequalities

\begin{equation}
S_{\mathrm{B}}>S_{2}>S_{3}>\cdot \cdot \cdot >S\left( D\right) .  \tag{12.2}
\end{equation}
For an independent-particle initial state, all these entropies are equal at
the initial time $t_{0}.$ As $S_{\mathrm{B}}$ does, any $S_{n}$ begins to be
equal to $S\left( D\right) $ before increasing so as to reach eventually $S_{%
\text{\textrm{eq}}}$ when global equilibrium is attained. The curves $%
S_{n}\left( t\right) $ are ordered according to (12.2). As $n$ increases,
the rise of $S_{n}$ takes place later and later. This is necessary because,
at any fixed time $t,S_{n}$ decreases with $n$ down to $S\left( D\right) ,$
which is attained when $n$ is so large that all the correlations created by
the evolution from $t_{0}$ to $t$ are taken into account. For $t$ and $n$
both large, $S_{n}\left( t\right) $ behaves \textit{non-uniformly:} it tends
to $S_{\text{\textrm{eq}}}$ when $t\rightarrow \infty $ for fixed $t.$ These
two regimes correspond to $t\gg n\tau $ and to $t\ll n\tau ,$ respectively,
where $\tau $ is the time between successive collisions, since each new
collision adds an additional correlation to a cluster already correlated by
the previous collisions.

Similar ideas apply to systems other than gases. In particular the adequacy
of a partial description based on some choice of the set $R$ relies on the 
\textit{absence of significant memory effects }in the resulting equations of
motion (8.11). If such effects occur, there exist within the set that we
regarded as irrelevant some observables which are coupled to those that we
included in $R$ and which have a comparable characteristic time. A decrease
of the relevant entropy indicates that information, or order, has been 
\textit{stored in such hidden variables, }and is \textit{released later on }%
towards the variables retained in our description. Any hidden variable of
this type should therefore be included into the relevant set $R,$ so as to
ensure that the information flowing towards the discarded variables is 
\textit{irretrievably lost.}

We illustrate this point by reconsidering the \textit{spin-echo experiment }%
described in the end of \S 3. Let us first forget about the experimental
possibility of rotating the spins by $\pi $ around the $x$-axis at the time $%
T$ by means of a pulse along this axis. The thermodynamics of spins is
characterized by an entropy $S_{\mathrm{th}}\left( M\right) $ which is the
relevant entropy associated with the macroscopic quantity, the magnetization 
$\mathbf{M.}$ It vanishes when all the spins are aligned, and increases as $%
M $ decreases. Thus the relaxation of $\mathbf{M},$ which can be described
by a macroscopic equation of motion involving both rotation and damping,
results as expected in the increase of $S_{\mathrm{th}}.$ However, the
magnetic pulse at the time $T$ along the $x$-axis not only changes $\mathbf{M%
}$ but also changes the individual spins in a manner which \textit{keeps
track of their history} between the time $0$ and the time $T.$ Owing to this
possibility of manipulation, the memory about the initial state is not lost
at the time $T$ although the spins point in any direction of the plane. This
apparent disorder of the spins reflects in fact the weak dispersion in the
values at each site of the permanent field along $z,$ since the spins have
rotated between the time $0$ and the time $T$ by an angle which is
proportional to the local field. Due to this correlation between spins and
fields, the total magnetization no longer follows between the times $T$ and $%
2T$ the macroscopic relaxation equation, but an equation involving memory
about the individual spin sites. The decrease between the times $T$ and $2T$
of the entropy $S_{\mathrm{th}}\left( M\right) $ down to its initial value
zero reflects the possibility of retrieval of the initial information
through this memory. We must thus include within the relevant set after the
time $T$ the hidden variables, i.e., the expectation values of the
individual spins instead of that of the total spin. This produces a set of
differential equations without memory, valid at all times. The relevant
entropy associated with the individual spins never decreases and is now
meaningful, contrary to the thermodynamic entropy $S_{\mathrm{th}}\left(
M\right) .$

The above review articles contain extensive bibliographies to which the
reader is referred.

\end{document}